# SINFONI - Integral Field Spectroscopy at 50 milli-arcsecond resolution with the ESO VLT

Frank Eisenhauer[a], Henri Bonnet[b], Roberto Abuter[a], Klaus Bickert[a], Fabio Bianca-Marchet[b], Joar Brynnel[b], Ralf Conzelmann[b], Bernard Delabre[b], Rob Donaldson[b], Jacopo Farinato[b], Enrico Fedrigo[b], Gert Finger[b], Reinhard Genzel[a], Norbert Hubin[b], Christof Iserlohe[a], Markus Kasper[b], Markus Kissler-Patig[b], Guy Monnet[b], Claudia Röhrle[a], Jürgen Schreiber[a], Stefan Ströbele[b], Matthias Tecza[a], Niranjan Thatte[a], and Harald Weisz[c]

[a]Max-Planck-Intitut für extraterrestrische Physik, Giessenbachstrasse, 85748 Garching, Germany
[b]European Southern Observatory, Karl-Schwarzschild-Straße 2, 85748 Garching, Germany
[c]Ingenierbüro für den Maschinenbau Weisz, Menzinger Straße 1, 80638 München, Germany

## ABSTRACT

SINFONI is an adaptive optics assisted near-infrared integral field spectrometer for the ESO VLT. The Adaptive Optics Module (built by the ESO Adaptive Optics Group) is a 60-elements curvature-sensor based system, designed for operations with natural or sodium laser guide stars. The near-infrared integral field spectrometer SPIFFI (built by the Infrared Group of MPE) provides simultaneous spectroscopy of 32 x 32 spatial pixels, and a spectral resolving power of up to 3300. The adaptive optics module is in the phase of integration; the spectrometer is presently tested in the laboratory. We provide an overview of the project, with particular emphasis on the problems encountered in designing and building an adaptive optics assisted spectrometer.

## 1. SINFONI: ADAPTIVE OPTICS AND INTEGRAL FIELD SPECTROSCOPY

SINFONI (**SIN**gle **F**aint **O**bject **N**ear-IR **I**nvestigation) is an adaptive optics assisted near infrared integral field spectrometer mounted to the European Southern Observatory (ESO) VLT (**V**ery **L**arge **T**elescope). The instrument is a combination of the Adaptive Optics module [1], a clone of MACAO (**M**ultiple **A**pplication **C**urvature **A**daptive **O**ptics), developed and built by ESO, and of the near infrared integral field spectrograph SPIFFI (**SP**ectrograph for **I**nfrared **F**aint **F**ield **I**maging) [2], developed and built by the Max-Planck-Institute for extraterrestrial Physics (MPE).

Currently, ESO offers two state-of-the-art near infrared instruments at the VLT: ISAAC [3] for seeing limited infrared imaging and spectroscopy, and NAOS/CONICA [4,5] for high order adaptive optics imaging and low-resolution spectroscopy. However, spectroscopy of faint objects with diffraction limited angular resolution at an eight-meter telescope will strongly benefit from a dedicated instrument, which combines the following characteristics: first, diffraction limited observations at near infrared wavelengths, optimized for faint wave-front reference stars and laser guide star operations; second, instantaneous spectroscopy of a two dimensional field with sufficiently high spectral resolution for deep observations between the night sky emission lines.

Both partner institutes collected extensive experience with diffraction-limited spectroscopy with their instruments ADONIS/SHARP [6] at the La Silla 3.6 m telescope, and ALFA/3D [7] at the Calar Alto Observatory 3.5 m telescope. Our conclusion is that when observing with adaptive optics, integral field spectroscopy gains significantly over long-slit spectroscopy and Fabry-Perot imaging. The latter suffers significantly from the variation of the sky emission and the point-spread-function (PSF) between consecutive images, and consumes exorbitant observing time for large wavelength coverage. Long-slit spectroscopy, on the other hand, lacks the essential two-dimensional information for decomposing the spatial flux distribution, and loses most of the source flux for a diffraction limited slit width and moderate correction of the atmospheric aberrations. In addition, flexure within the instruments complicates the acquisition of objects, and it is difficult to keep the object in the slit for long integration times. After an initial negotiation process in 1997, the actual development of SINFONI started in late 1999 with the preliminary design review of SPIFFI. Despite the fact that the instrument is entering virgin foil on many fronts, the whole project was planned to be fast track with first light at the telescope in 2004.

The Adaptive Optics Module of SINFONI is based on MACAO [8,9], developed for the VLT interferometer. Our intention is to clone the corresponding subunits of the MACAO wavefront sensor wherever possible, and to develop a similar corrective assembly (deformable mirror and the tip/tilt mount). There are several advantages associated with this approach: we share the development effort of key components in house, and we can concentrate on the SINFONI specific aspects. However, there are several fundamental differences to the standard MACAO system: The instrument is mounted at the Cassegrain focus, and thus exposed to varying gravity loads; and the system can operate with a laser guide star combined with a natural tip/tilt reference. The working principle of the integral field spectrometer SPIFFI is copied from its precursor 3D [10]. Both instruments are built around an image slicer with plane mirrors [11]. However, SPIFFI is fully cryogenic, including the image slicer, and the number of spatial pixels is four times as large. With four different gratings the spectrometer delivers optimum efficiency over the entire 1.05 - 2.45 µm wavelength range, and three different image scales allow observations both with seeing limited and diffraction limited angular resolution. The combination of the adaptive optics module and the spectrometer fits within the size and weight constraints of a Cassegrain instrument at the VLT. Figure 1 shows a computer rendered picture of SINFONI mounted to the telescope.

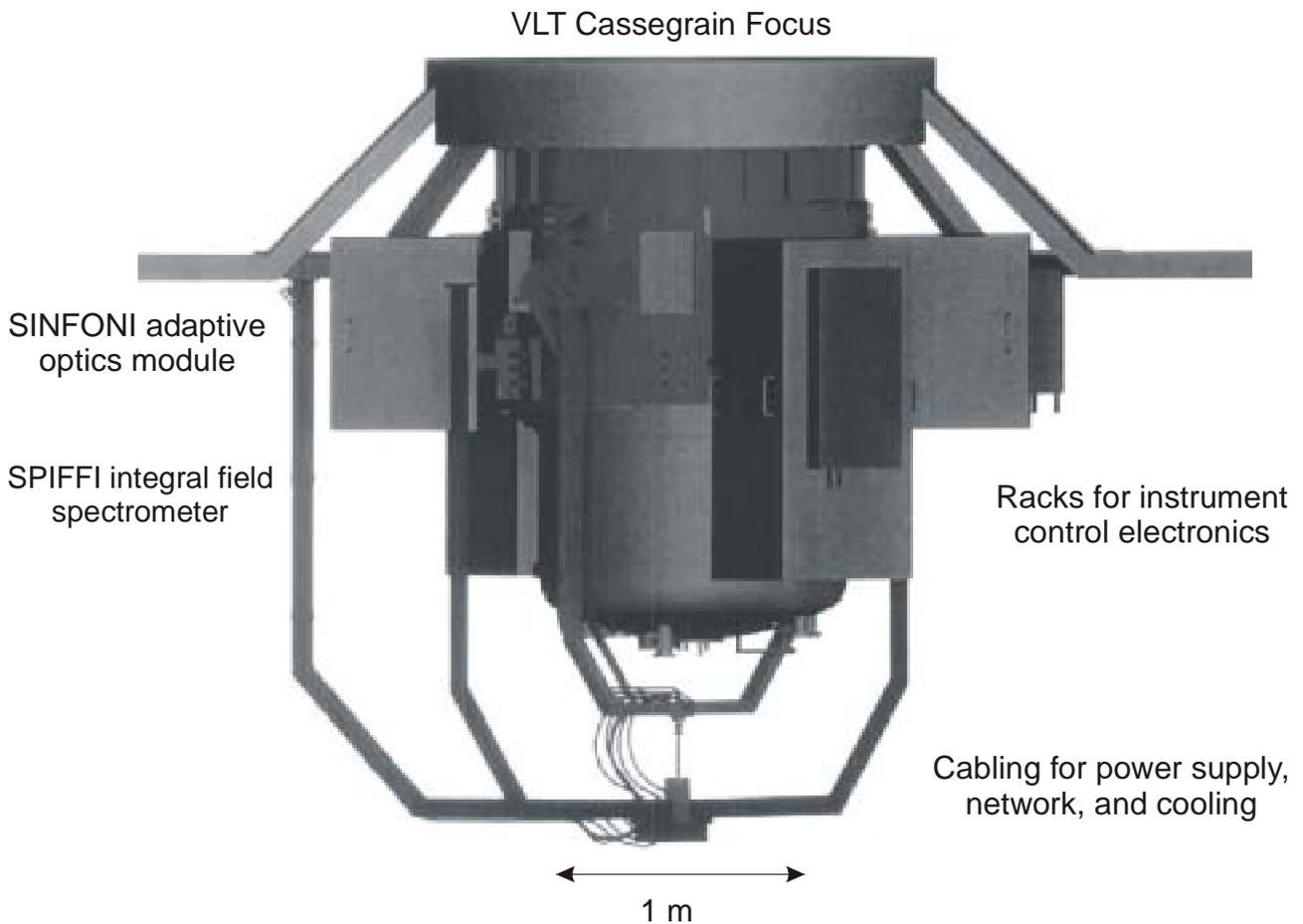

*Figure 1. Computer rendered picture of SINFONI: The upper cylinder hosts the adaptive optics module, the cryostat of the integral field spectrometer is mounted below.*

## 2. ADAPTIVE OPTICS MODULE

The SINFONI Adaptive Optics (AO) Module, designed and built by ESO, images the pupil onto a 60-element bimorph deformable mirror (DM). The corrected reflected beam is then re-imaged to the SPIFFI focus, located after the infrared dichroic. The science beam (1.05 - 2.45 µm) is transmitted to SPIFFI, and the visible light (450 – 950 nm) is reflected towards the wavefront sensor optical path. A field selector picks up the reference star in an intermediate image plane in the 1x2 arcmin$^2$ total field of view. The beam is then sent to the curvature wavefront sensor, fully recurrent from the MACAO program: an image of the reference star is formed onto the membrane mirror, and the beam is collimated by a beam expander, creating a pupil image on the lenslet array. The beam intercepted by each of the 60 sub-apertures is then fed to an optical fiber, connected to an Avalanche Photo Diode (APD), located in a cabinet outside the AO-Module. The membrane mirror is acoustically excited at 2.1 kHz, enabling the intra- and extra-pupil areas to be imaged successively on the lenslet array. The APD counters are read out synchronously with the membrane oscillations and the relative flux difference between the intra- and extra-pupil integration provides the curvature signal. The Real Time Computer (RTC) filters this signal and projects it in mirror space to generate an update of the DM command at a rate up to 500 Hz.

AO corrections require the availability of a bright wavefront reference near the Line of Sight (LOS) of the science field. It is well known that bright reference stars suitable for AO correction is scarce. The field of view within the isoplanatic angle at the observing wavelength considered is also limited to few tens arcsec. The net result is the relatively low sky coverage of a natural guide star AO system. In order to increase the sky coverage the AO-Module is also designed to use a laser guides star (LGS) as a wavefront reference [12]. For LGS operations, a LGS dichroic is inserted in the optical path between the IR dichroic and the field selector. The LGS dichroic transmits the sodium light at 589 nm towards the Curvature Wavefront Sensor (CWFS) for high order wavefront sensing and reflects the visible light of the Natural Guide Star to the tip-tilt sensor (STRAP). The tip-tilt star is selected within the 1x2 arcmin$^2$ FOV and imaged onto the STRAP head. The RTC combines the STRAP tip/tilt signal with the high order error signal to generate a DM command update, while the tip/tilt modes from the CWFS are sent to the Laser Guide Star Facility (LGSF) to feed the LGS pointing control loop. A detailed description of the instrument design can be found in [1]. In this section we will concentrate on the problems encountered in the design and manufacturing of the instrument.

### 2.1. Deformable mirror

The deformable mirror is based on a Bimorph Piezo Electric Transducer (PZT) principle, developed successfully by CILAS since 10 years for many application (François Roddier system was the first example in 1991): 60 electrodes are sandwiched in-between two PZT plates enabling the PZT material to be polarized in opposite directions. This produces a curvature over the electrode surface, proportional to the applied voltage.

The stroke needed on each electrode to correct for the atmospheric turbulence is proportional to the atmospheric focus variance over the size of one sub-aperture. For MACAO, the size of the pupil image on the DM is 60 mm, and there are 60 electrodes leading to ~ 8 mm sub-apertures, i.e. ~ 40% smaller than in previous applications. This requires a minimum radius of curvature twice smaller than achieved with state-of-the-art Bimorph mirror.

There are two ways to reduce the minimum radius of curvature: either by increasing the control voltage range or by reducing the PZT plate or glass layer thickness. The first solution was rejected for safety reasons as well as standardization of the MACAO DM control electronics. The capability to reduce the DM thickness was demonstrated on a prototype by CILAS, and this solution was selected as a baseline.

The serial units all meet the requirements in terms of stroke while the voltage required to flatten the mirror does not exceed 10% of the available range. The remaining concern is the increased sensitivity of the mirror to changes in the gravity load, which may limit the exposure time in open loop mode. This could be cured by improving the DM mounting, as currently explored conjointly with the manufacturer.

### 2.2. Image motion minimization

The original design of the spectrograph is based on a $1024^2$ detector (see §3.7). In this design, the number of pixels is too small to Nyquist-sample the spectra in a single exposure. Instead, we recover the full spectral resolution by shifting the spectra by half a pixel in a second exposure. This spectral dithering technique requires that the image does not move more than 1/5 pixel between the two exposures. For integration times of 30 minutes, this would lead to an acceptable image motion rate of 10 milliarcseconds (mas) per hour.

There are several independent contributors to the image motion budget: (a) Telescope guiding errors about the line of sight, combined with the lever arm between the tip/tilt reference and the science field of view. This is a minor contribution to the image motion budget, evaluated to ~ 2 mas/hr for guiding star up to 20 arcsec off-axis. (b) Differential atmospheric refraction between the tip/tilt sensor guiding wavelength and the science wavelength combined with the error in the estimation of the guiding wavelength. This effect accounts for 20 mas/hr for observation down to 45° from zenith. (c) Changes in the atmospheric dispersion due to air-mass variations corresponding to an image motion of ~ 5 mas/hr. (d) Instrument's flexure along non-common optical paths of 25 mas/hr in NGS mode, and 35 mas/hr in LGS mode. We are confident that 75% of this contribution will be removed through an adequate pointing model correcting in open loop the flexure induced pointing error according to the current orientation of the gravity load.

Combining these contributions quadratically, we find an image motion rate of ~ 20 mas/hr, i.e. twice the initial requirement. Since the dominant contributor to the image motion budget is the guiding wavelength error, it is clear that the 10 mas/hr requirement can not be achieved even with an infinitely stiff instrument. The impact would have been a reduction of the maximum exposure time, which would reduce the performance of the instrument in the high spatial resolution mode (25 mas/pixel), which is limited by the detector readout noise.

In addition to the scientific drivers outlined in [14], the image motion requirement and the difficulties in the manufacturing of a fast f/D = 1.45 camera (see §3.7) motivated the upgrade of the instrument with a $2048^2$ Hawaii II detector, which will enable the full spectral resolution to be achieved in one exposure. With the 2k camera, the image motion requirement is derived from the image quality need: our goal is to not degrade the Strehl by more than 20% in a single exposure. The limitation to the exposure time arising from the image motion is then ~ 90 minutes in J band and almost 3 hours in K band.

### 2.3. Lenslet array

The lenslet array intercepts the beam in a pupil plane and splits the pupil in 60 sub-apertures. Each sub-pupil is re-imaged on the core of a 100 µm fibre, which feeds the light to an Avalanche-Photo-Diode. The geometry of the sub-apertures (see *Figure 2*), the sag of the lenslet, the encircled energy and the accuracy of the lenslet vertex altogether constituted a very demanding design. The 60 apertures keystone geometry is distributed in five rings of varying number of lenslets: 4, 8, 12, 16 and 20. In order to reduce the required sag, two lenslets are used: a first one (*Figure 2*) made of 2 keystones shape apertures back to back; this allows to realise the 45 mm focal length needed with less sag on each lenslet. The second one is made of 60, 800 µm circular lenses deposited on a substrate and located at the vertex of the first lenslets. It is used to inject the light into the optical fibres.

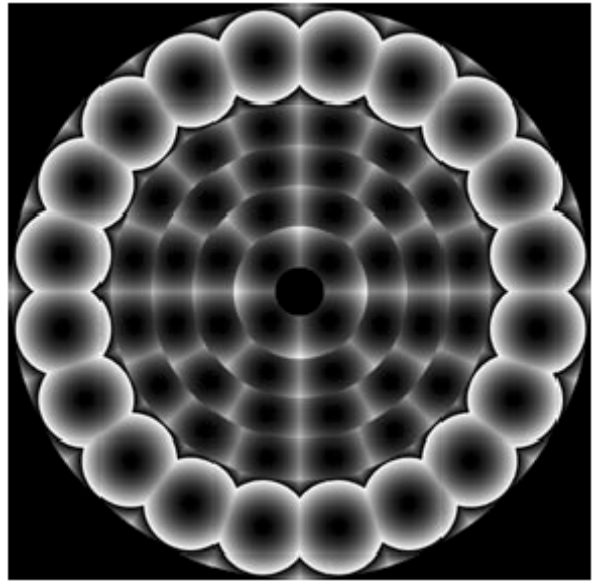

*Figure 2. Geometry of the lenslet array: Two such arrays are glued back to back in order to produce the first lenslet. The gray level indicates the sag of each lenslet; the external ring is made of fresnel lenses in order to reduce the sag needed. A master is fabricated with a laser writing technique and replicated to produce the deliverable units (Heptagon, Zurich, Switzerland).*

### 2.4. Focus Control in LGS mode

In LGS mode, the primary AO loop is fed by the CWFS to control the focus. The CWFS must therefore be focused to the distance of the LGS, which varies with: (a) the altitude of the mesospheric sodium layer (range ~ 80 - 110 km); (b) $(\cos \Psi)^{-1}$, where $\Psi$ is the zenith angle (range 0 - 60 deg). The focus error induced by an error in the estimation of the Sodium Centroid Altitude (SCA) is:

$$\delta = \cos \Psi \left(\frac{Df}{H}\right)^2 \delta H$$

i.e. $\delta \sim 0.2$ mm of defocus for $\delta H = 100$ m at zenith, and twice less at 60° from zenith. The corresponding wavefront error (fourth Zernike coefficient) writes:

$$z_4 = \frac{\cos \Psi}{16\sqrt{3}} \left(\frac{D}{H}\right)^2 \delta H$$

i.e. $z_4 \sim 25$ nm (Strehl = 99.5%) for $\delta H = 100$ m at zenith.

Our current evaluation of the short term variability of the SCA correspond to a typical Strehl degradation factor of 0.8 in K for 1 hr exposures if the trombone position is tuned to the initial SCA but not adjusted in closed loop during the observation.

Such a closed loop control is difficult to implement, because in LGS mode, the Curvature Wavefront Sensor is equally sensitive to (a) a defocus induced by the atmosphere or the telescope (also affecting the science beam), and (b) a defocus induced by a SCA variation (not affecting the science beam). The timescales of these two contributions are not sufficiently separated to allow disentangle them efficiently by spectral filtering. The baseline control strategy we selected is illustrated in *Figure 3*.

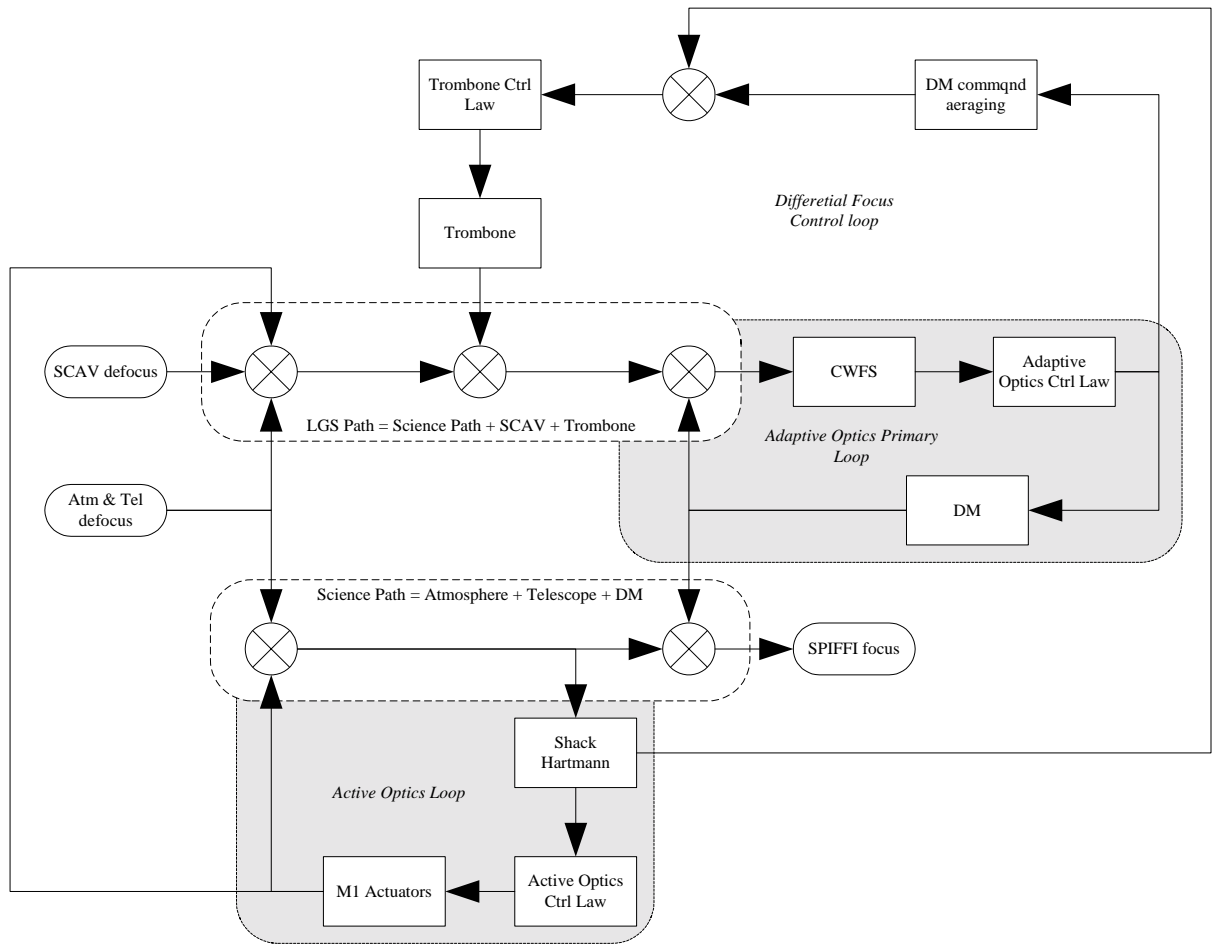

*Figure 3. Baseline trombone position control strategy*

The Shack Hartman of the VLT Active Optics System senses a Natural Guide Star, which is affected by the atmosphere and the telescope defocus. Meanwhile, the Curvature Wavefront Sensor of the Adaptive Optics senses the Laser Guide Star, which is affected, beside the atmosphere and the telescope defocus, by the SCA variation defocus and by the trombone position error. The difference between the Shack Hartmann measurements and the Adaptive Optics Deformable Mirror command averaged over the Shack Hartmann integration time provides therefore an estimation of the trombone position error.

In addition to this closed loop control, an open loop feed forward correction is fed to the trombone position to compensate for the focus variation expected during the next control cycle from geometrical variation of the Laser Guide Star distance due to changes in the zenith angle.

The error budget on the defocus due to trombone mis-positioning is dominated by the loop error (25 nm) and by the feed forward error (16 nm) leading to a total rms error of 35 nm.

# 3. INTEGRAL FIELD SPECTROMETER SPIFFI

The integral field spectrometer SPIFFI provides simultaneous spectroscopy of 32 x 32 spatial pixels with a spectral resolving power of approximately 3300. When used with the adaptive optics – the angular resolution at a wavelength of 2.2 µm is 56 mas (milli-arceconds) - the size of a pixel corresponds to 25 mas. To also make optimal use of the spatial elements in seeing limited observations - and at partial correction of the atmospheric turbulence with faint reference stars - SPIFFI provides image scales of 100 mas / pixel and 250 mas / pixel. Each of the atmospheric transmission bands J (1.1 – 1.4 µm), H (1.45 – 1.85 µm), and K (1.95 – 2.45 µm) are covered with a single exposure. In addition, the combined H and K bands (1.45 – 2.45 µm) can be observed at correspondingly lower resolution. With the help of additional mirrors, the observer can observe simultaneously the night sky emission from a blank field up to 45 arcseconds off-axis. SPIFFI is already fully assembled, and at the moment undergoing extensive test in the laboratory. A detailed description of the instrument design can be found in [2]. In this section we will concentrate on the problems encountered in the design and manufacture of the instrument.

## 3.1. Cryostat

*Figure 4* shows the opto-mechanical components of SPIFFI. All components are cooled in a bath cryostat to the temperature of liquid nitrogen. The liquid nitrogen reservoir sits below the instrument plate. One potential problem in the design of light-weighted bath cryostat is the selection of the proper Aluminum alloy. The thermal conductivity of these alloys varies strongly at the temperature of liquid nitrogen. For optimum cooling we would have preferred Al 6061, an Aluminum alloy with a significant content of Silicon, but this alloy is not included in the list of materials for pressure vessels suggested by the German norms and technical recommendations (VdTÜV). Also to ease welding of the liquid nitrogen tank, we had to revert to the Aluminum Magnesium Manganese alloy Al 5083.

## 3.2. Sky-spider

The light enters SPIFFI from the top, and first passes the so-called sky-spider. This device contains three motorized pairs of mirrors, which reflect the light from an off axis sky field onto the image slicer field of view. The projected distances of the sky-positions from the field center are 15, 30, and 45 arc-seconds. A forth position can be used for an upgrade with a coronographic mask.

## 3.3. Pre-optics

Below the sky-spider, a doublet lens unit collimates the light onto a cold stop for the suppression of the thermal background. The stop has a diameter of 6 mm, and is equipped with a central obscuration with the equivalent size of the telescope secondary mirror. Just in front of the cold stop is the motorized filter wheel, housing the four band-pass filters, which are required to suppress the unwanted diffraction orders of the gratings. After the cold stop, the motorized optics wheel provides the interchangeable lens systems for the three different image scales, and a pupil-imaging lens for alignment purposes. Each objective consists of two or three lenses. All moving functions are driven by five phase stepper motors, which have been modified for use at cryogenic temperatures. Because it has been observed that some of these modified motors malfunctioned after some time, we equipped the motors with encoders. In three months of laboratory operation, we have not experienced any problem with the motors, but one encoder failed after a couple of cool-downs.

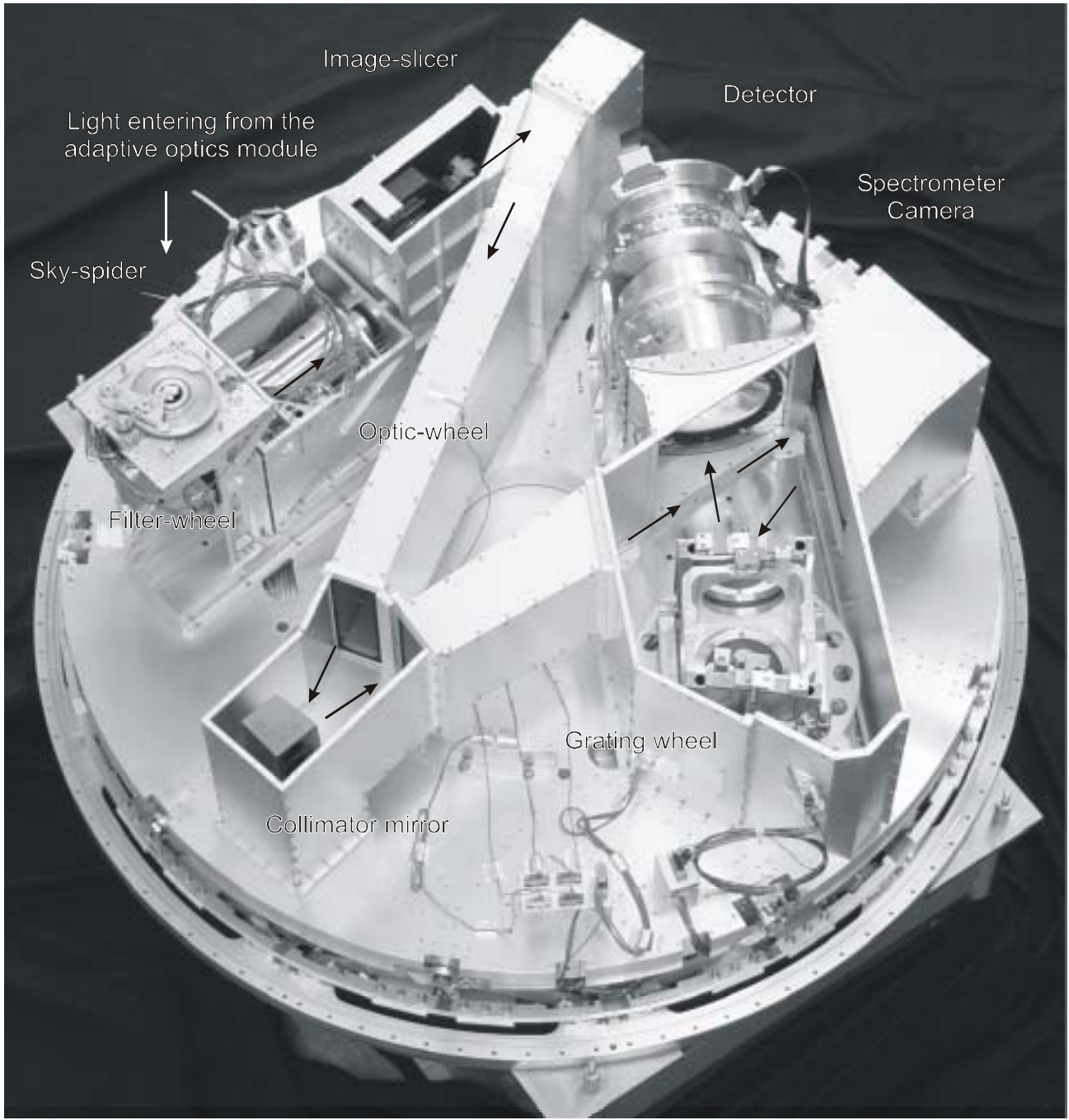

*Figure 4.* An inside view of SPIFFI: The cryostat cover and the reinforcing structure have been removed to provide a free view on the opto-mechanical components of SPIFFI. The light enters from the top, and passes the so-called sky-spider. The pre-optics with a filter-wheel and interchangeable lenses provides three different image scales. The image slicer rearranges the two-dimensional field in a pseudo-long slit, which is perpendicular to the base plate. Three diamond turned mirrors collimate the light on the gratings. In total, four gratings are implemented on the grating drive. A multiple-lens system then focuses the spectra on a Rockwell HAWAII array. The diameter of the instrument is 1.3 m.

### 3.4. Image slicer

The focus of the pre-optics is located at the so-called small slicer. This part of the image slicer consists of a stack of 32 plane mirrors, which slices the image into slitlets, and separates the light from each slitlet in different directions. A second set of 32 mirrors, the so-called big slicer, collects the light and forms the pseudo-slit. *Figure 5* shows an image of the image slicer. To avoid differential thermal contraction, the unit is completely made from a zero expansion glass. All parts (approximately 70 in number) are optically contacted. No glue is used in the assembly. A detailed description of the design and the manufacturing technique can be found in [11,13]. The problems in the manufacturing of the slicer are mainly in the process of cleaning and handling. Specific care has to be taken in the cleaning of the small slicer mirrors, which have a width of only 300 µm. In a first attempt, these mirrors have been cleaned with cotton swabs and commercial cleaning fluid, but because of the small area, the cleaning caused significant scratches with a depth of approx. 100 nm. The small inset of *Figure 5* shows the interferogramm of a small slicer mirror with the described scratches, which can be seen as jumps in the fringe pattern.

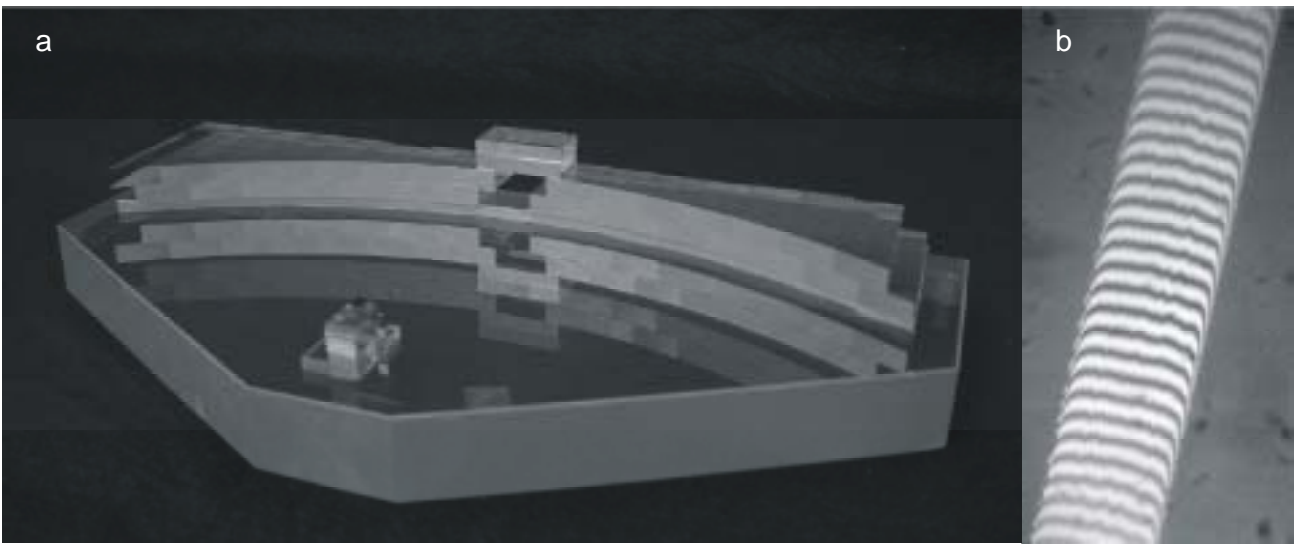

*Figure 5.* SPIFFI image slicer (a): The light is entering through the hole in the big slicer. A stack of 32 small mirrors – the so-called small slicer – slices the image and redirects the light towards the 32 mirrors of the big slicer, which rearranges the slitlets to a 31 cm long pseudo-slit. The small picture (b) shows an interferogramm of the 300 µm wide mirror of the small slicer. The jumps in the fringe pattern indicate scratches, which have been present in the prototype, but can be avoided by less aggressive cleaning techniques.

### 3.5. Spectrometer collimator

After the image has been sliced and rearranged to the pseudo slit, three diamond turned mirrors collimate the light onto the gratings. The first mirror is spherical, and the other two mirrors have an oblate elliptical shape. The elliptical mirrors are shown in *Figure 6*. All mirrors are made from Aluminum, with a Nickel layer for the diamond turning process. The mirrors are gold-coated for higher reflectivity. Despite the fact that the mirrors have been post polished after the diamond turning process, turning marks are clearly visible. These turning marks are presently limiting the performance of the collimator. *Figure 6* shows the interferogramm of the collimator in on-axis operation. The turning marks of all three mirrors are clearly visible, distinguishable by the different centers of rotation. The dominant turning marks have a typical depth of 200 nm, and a typical width of one millimeter. The patchy appearance of the interferogramm results from the combination of individual measurements to fill the whole pupil, and is not present in the actual wavefront.

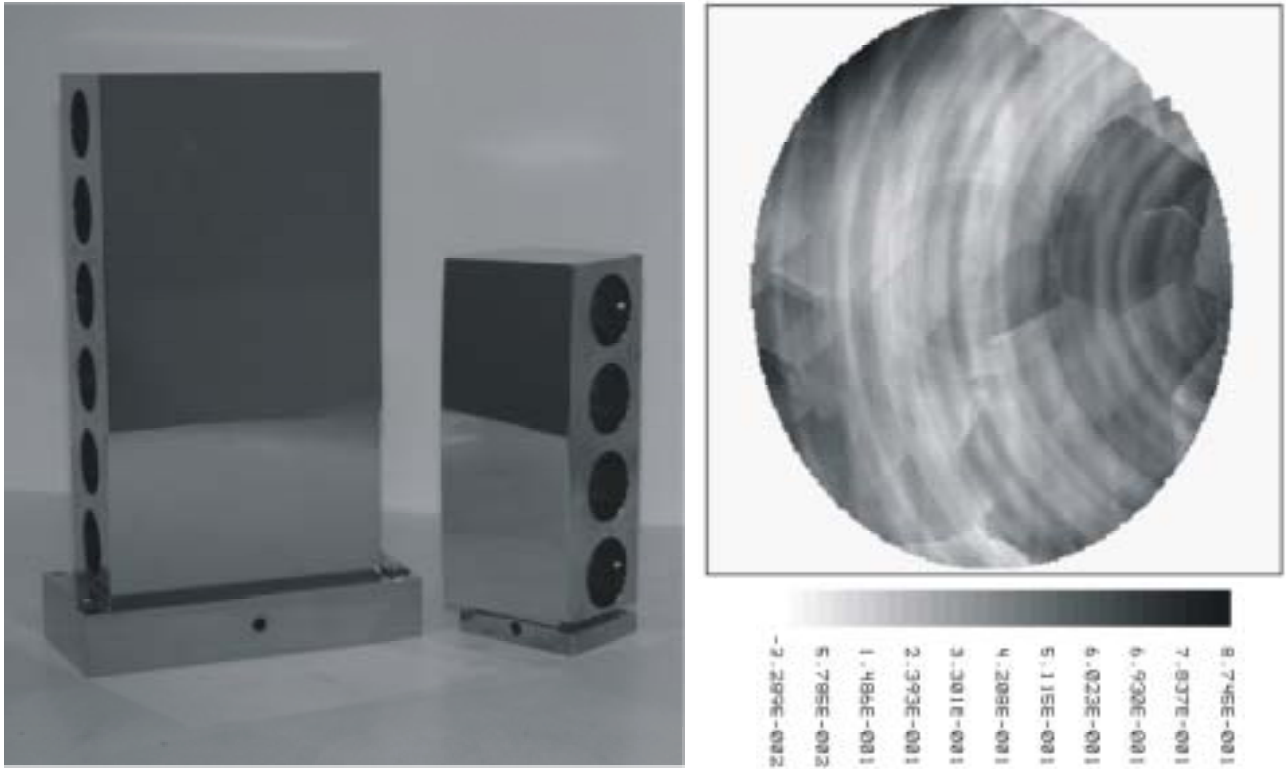

*Figure 6. Collimator mirrors: The mirrors have been diamond turned in Nickel coated Aluminum. Despite the post-polishing, all mirrors show turning marks, which presently limit the performance of the collimator. The interferogramm shows the on-axis wavefront error in single pass. The unit is waves at 1.1 μm.*

### 3.6. Grating wheel

In total, SPIFFI has four gratings to cover the three atmospheric J, H, and K windows, and the combined H&K band. The gratings are directly ruled in gold on an Aluminum substrate, and every grating is blazed to the center of its pass-band. In order to be able to move the spectrum on the detector with an accuracy of 1/5 of a pixel, the grating wheel is driven with a 1:647 gear, and controlled via an inductive encoder directly attached to the grating wheel. Both devices have been working without problems so far.

### 3.7. Spectrometer camera

In its final configuration [14], SPIFFI will be equipped with a Rockwell 2k x 2k pixel Hawaii II array, and a lens spectrometer camera. A NOVA/MPE/ESO consortium will jointly build this camera. Until the delivery of this detector, SPIFFI will be operated as a VLT guest instrument with a 1k x 1k pixel Hawaii1 detector [15]. This configuration has the disadvantage that the spectrometer camera has to be operated at a very low f-number of $f/D = 1.45$. The 1k-camera (*Figure 7*) is a six-lens system made from Barium Fluoride and Schott IRG2 glass. The diameter of the three big lenses is 160 mm. While we had no problems with the procurement and manufacturing of the IRG2 lenses, Barium Fluoride lenses of this diameter turned out to be a challenge on its own. The problems started with the difficulties to procure mono-crystal material with sufficiently low refractive index inhomogenity and birefringence, and continued with damages during polishing and coating. In summary, the manufacturing of six large blanks extended over two years. Because of its very fragile structure and soft surface, the manufacturing of approximately every second large lens failed, one lens broke during coating process (*Figure 7*c). All lenses have a multi-layer antireflection coating optimized for the wavelength range from 1.05 – 2.45 µm. The oxide-based coating has been applied in an ion-assisted deposition process. Because of the slight hygroscopic nature of Barium Fluoride, the lenses have to be cleaned carefully immediately after the polishing, and should not be exposed to ambient air for long times. One of our coating runs of the pre-optics lenses failed completely. *Figure 7*d shows an example of a Barium Fluoride lens where the coating is peeling off.

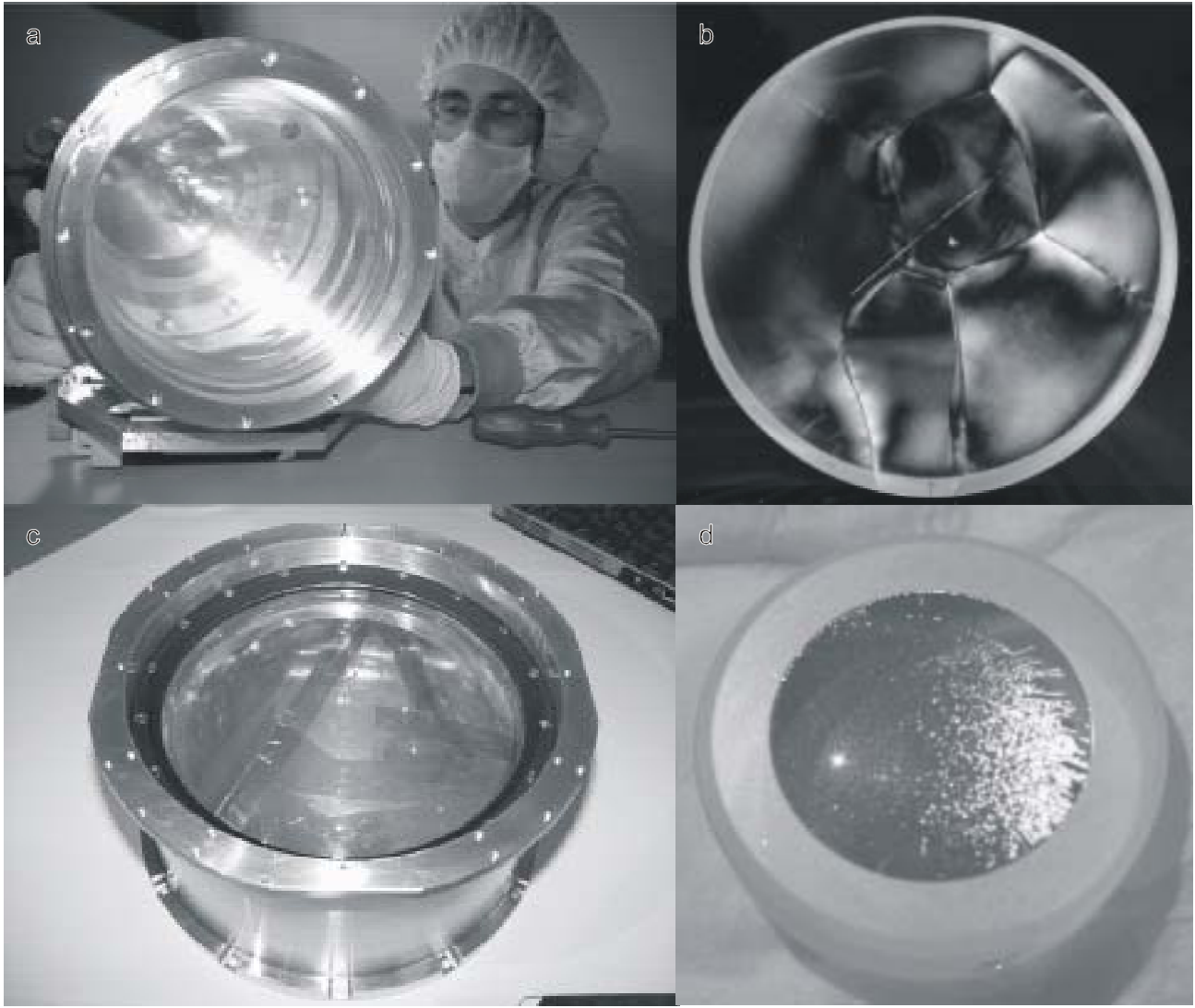

*Figure 7.* SPIFFI 1k camera (a): The f-number of this six-lens camera is f/D = 1.45, the diameter of the large lenses is 160 mm. The lenses are made from Barium Fluoride and Schott IRG2 glass. Specifically the Barium Fluoride lenses have been very difficult to manufacture: The manufacturing process of six large Barium Fluoride blanks took two years, because of the unavailability of mono-crystal material. Figure b shows the hetero-crystalline structure of a rejected crystal, which is placed between crossed polarizers to measure birefringence. Barium Fluoride is rather fragile and soft. Approximately half of the large crystals have been damaged in the manufacturing process. Figure c shows an example of a big lens, which broke in the coating process, and which was glued together for the very first cool-down of SPIFFI. Because of the slight hygroscopic nature of Barium Fluoride, special care must be taken in the cleaning, and exposing to humidity between the polishing and coating process should be avoided. Figure d shows a pre-optic lens where the coating is peeling off.

Because of the fast f/D ~ 1.45 beam in the SPIFFI 1k camera, the detector must be focused with an accuracy of better than approximately 1/100 mm. This is achieved with three micrometer-screws. However, given the large mass of several hundreds of kilograms of the cold structure, and the large vacuum volume of approximately one cubic meter, every warm-up and cool-down cycle including evacuation takes typically one week. Therefore we are presently exploring the implementation of piezo-driven pico-motors instead of the micrometer-screws. Unfortunately, the torque of these motors is strongly reduced at cryogenic temperature, and we have not been able to operate the motors reliably at a temperature of 77 K. Instead we are implementing a small heater with every pico-motor, to heat the device temporarily for the focusing process.

## 4. SYSTEM PERFORMANCE

### 4.1. Adaptive optics module

The SINFONI adaptive optics module is particularly optimized for observations of faint objects. The use of photon-counting avalanche photodiodes, and the corresponding low order correction pushes the performance to fainter magnitudes when compared to the NAOS adaptive optics. But this gain is achieved at the cost of a reduced peak performance at shorter wavelengths. In addition, the pure tip/tilt correction with STRAP will allow partial correction for even fainter targets. *Table 1* summarizes the predicted performance for typical observations with a natural guide star. The details of the simulations are described in [16]. When the Laser Guide Star Facility (LGSF) [17,18] is installed at the VLT, SINFONI will also support observations with the artificial guide star. The predicted performance in this mode is summarized in [1].

|        | 10 mag | 12 mag | 14 mag | 16 mag | 18 mag |
|--------|--------|--------|--------|--------|--------|
| K-Band | 58%    | 56%    | 44%    | 22%    | 5%     |
| H-Band | 38%    | 35%    | 24%    | 7%     | -      |
| J-Band | 19%    | 16%    | 8%     | -      | -      |

*Table 1. Predicted performance of the SINFONI adaptive optics with a natural guide star. The table contains the predicted Strehl number in the three atmospheric bands J, H, and K for given brightness of the wavefront reference star and a visible seeing of 0.65 arcsec.*

### 4.2. Spectrometer

The sensitivity of the spectrometer is summarized in *Table 2*. Specifically for faint objects with partial correction by the adaptive optics, SPIFFI will still collect the total flux when classical spectrometers will lose most of the photons outside a diffraction limited long-slit. But also seeing limited observations will gain from the integral field capability. The details of the sensitivity calculations are described in [19]. *Figure 8* illustrates the raw data format of SINFONI. The figure shows the detector illumination for the first laboratory observations of a Xenon and Neon lamp in the H-Band. A dedicated data reduction pipeline will calibrate the data and rearranges the measurement in a three-dimensional data cube with two spatial and one spectral dimension.

|        | Adaptive optics observations and 25 mas / pixel image scale | Seeing limited observations and 250 mas / pixel image scale |
|--------|-------------------------------------------------------------|-------------------------------------------------------------|
| J-Band | 19.2                                                        | 20.4                                                        |
| H-Band | 18.6                                                        | 20.1                                                        |
| K-Band | 18.0                                                        | 18.0                                                        |

*Table 2. Predicted Sensitivity of SPIFFI: The table lists the limiting magnitudes of the spectrometer for point-sources. The numbers are calculated for a one hour observation on source, and a signal-to-noise ratio of 5 per spectral resolution element between the night sky-emission lines.*

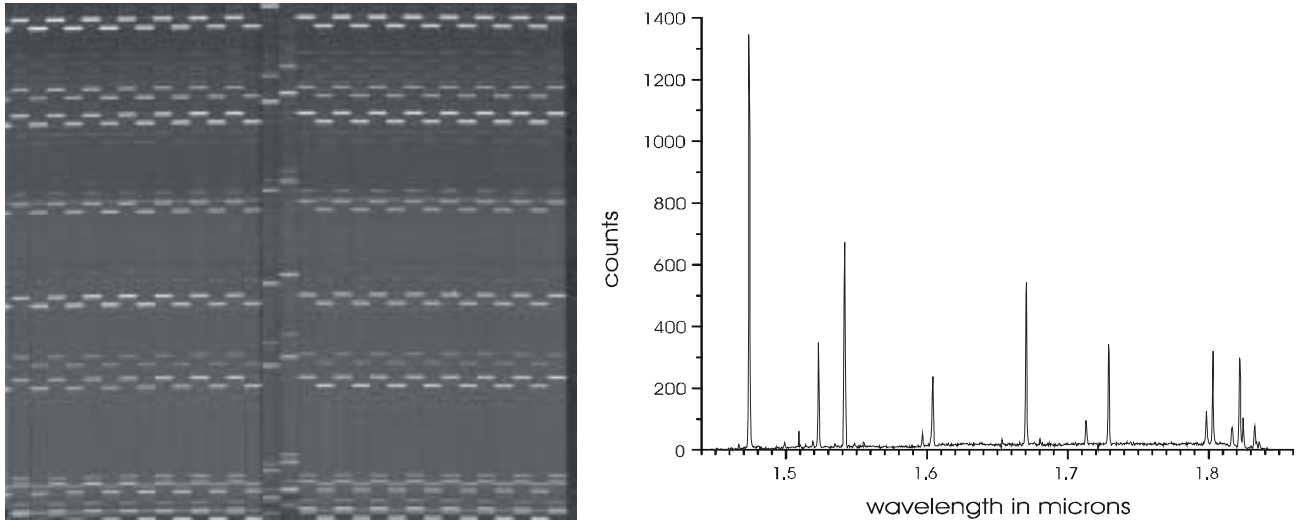

*Figure 8.* Raw detector frame and spectrum of SINFONI: The left figure shows the illumination of the detector for an observation of a Neon and Xenon calibration lamp. Clearly visible is the brick-wall pattern of the pseudo slit from the image slicer. The right graph shows a vertical cut along the dispersion direction.

## 5.  STATUS AND SCHEDULE

At the time of this conference (August 2002), the integral field spectrometer has been fully integrated with its 1k camera, and is undergoing extensive system level tests and debugging in the laboratory at MPE. All major subunits of the Adaptive Optics Module have been delivered, and are presently tested individually at ESO. We expect the first adaptive optics loop closed in the laboratory in October. After the flexure tests at the telescope simulator, SPIFFI will be shipped to Chile over New Year, and will be installed at the VLT for seeing limited observations without the adaptive optics. In parallel, the adaptive optics module will be tested with the telescope simulator in Garching, and the NOVA/ESO/MPE consortium will built the 2k camera the for the upgrade of SPIFFI. In mid 2003, we will combine the SINFONI adaptive optics module and spectrometer for system tests at the telescope simulator. The installation at the telescope is foreseen in early 2004. Laser guide star operation is foreseen for late 2004.